**Towards Space Group Determination from EBSD Patterns: The Role of Deep Learning and High-throughput Dynamical Simulations**

Alfred Yan[1], Muhammad Nur Talha Kilic[2], Gert Nolze[3], Ankit Agrawal[2,4], Alok Choudhary[2], Roberto dos Reis[1,4,5], Vinayak Dravid[1,4,5*]

[1] Department of Materials Science and Engineering, Northwestern University, Evanston, IL, United States.
[2] Department of Electrical and Computer Engineering, Northwestern University, Evanston, IL, United States
[3] Federal Institute for Materials Research and Testing, Berlin, Germany
[4] International Institute of Nanotechnology, Northwestern University, Evanston, IL, United States
[5] The NUANCE Center, Northwestern University, Evanston, IL, United States

* Corresponding author: v-dravid@northwestern.edu

**Funding Statement:** This research was supported in part through the computational resources and staff contributions provided for the Quest high performance computing facility at Northwestern University which is jointly supported by the Office of the Provost, the Office for Research, and Northwestern University Information Technology. This work made use of Northwestern University's NUANCE Center, which has received support from the SHyNE Resource (NSF ECCS-2025633), the IIN, and Northwestern's MRSEC program (NSF DMR-2308691). The authors also acknowledge funding from the International Institute for Nanotechnology at Northwestern University, and the Predictive Science and Engineering Design cluster program supported by the Northwestern Interdisciplinary Cluster Initiative.

**Conflict of Interests:** A.Y., R. R, and V. D. are authors on a provisional patent application based on the methods presented in this work.

**Keywords:** deep learning, EBSD, simulation, symmetry, space group, domain adaptation, electron microscopy

**Abstract:**

The design of novel materials hinges on the understanding of structure-property relationships. However, in recent times, our capability to synthesize a large number of materials has outpaced our speed at characterizing them. While the overall chemical constituents can be readily known during synthesis, the structural evolution and characterization of newly synthesized samples remains a bottleneck for the ultimate goal of high throughput nanomaterials discovery. Thus, scalable methods for crystal symmetry determination that can analyze a large volume of material samples within a short time-frame are especially needed. Kikuchi diffraction in the SEM is a promising technique for this due to its sensitivity to dynamical scattering, which may provide information beyond just the seven crystal systems and fourteen Bravais lattices. After diffraction patterns are collected from material samples, deep learning methods may be able to classify the space group symmetries using the patterns as input, which paired with the elemental composition, would help enable the determination of the crystal structure. To investigate the feasibility of this solution, neural networks were trained to predict the space group type of background corrected EBSD patterns. Our networks were first trained and tested on an artificial dataset of EBSD patterns of 5,148 different cubic phases, created through physics-based dynamical simulations. Next, Maximum Classifier Discrepancy, an unsupervised deep learning-based domain adaptation method, was utilized to train neural networks to make predictions for experimental EBSD patterns. We introduce a relabeling scheme, which enables our models to achieve accuracy scores higher than 90% on simulated and experimental data, suggesting that neural networks are capable of making predictions of crystal symmetry from an EBSD pattern.

**Introduction**

Global societal challenges require the development of new materials, particularly crystalline materials for widespread applications like photonics, electronics, and catalysis. For example, in the field of energy and sustainability, the development of novel catalysts can accelerate society's transition to renewable energy sources and alleviate the reliance on expensive rare metals used in present catalysts. High-throughput synthesis techniques are especially promising for discovering novel crystalline materials, with advances in megalibraries enabling the synthesis and screening of large collections of diverse crystalline nanomaterials for properties of interest [1–3]. However, to understand how functional properties relate to the compositions and crystal structures present in the megalibrary, the development of high-throughput characterization techniques that can determine the morphologies, compositions, and crystal structures of the large volume of different material phases in a short amount of time are needed.

Electron diffraction is a promising tool [4–7] for crystal structure determination, since it can be used to classify the crystal symmetry of a phase and thus narrow down possible structure prototypes. Deep learning methods hold potential for this, and methods have been applied in electron diffraction modalities like selected area electron diffraction (SAED) and convergent beam electron diffraction (CBED) [8–14]. There has also been related research in X-ray diffraction (XRD) [15–20].

Electron Backscattered Diffraction (EBSD) is an established SEM diffraction technique[21] that can be a promising alternative method for high-throughput symmetry determination. In EBSD, a focused electron beam hits a crystalline sample, and the backscattered electrons experience Bragg diffraction before colliding with a flat detector. The multiple scattering phenomena in EBSD causes the electron signal to appear in the form of Kikuchi bands and yield information about the 3-D crystal structure. For example, unlike other electron diffraction modalities, multiple low-indexed zone axes could be visible in a single pattern in the form of band intersections. Background correction of the signal results in a Backscatter Kikuchi Diffraction (BKD) pattern, also colloquially known as an EBSD pattern. Another advantage of using an SEM is the relatively large sample chamber, which is conducive to the analysis of larger samples that are characteristic of megalibrary chips.

Present applications of EBSD include orientation mapping and phase differentiation [22–28], including determining chirality [29,30], polarity [31,32], and tetragonality [27]. Many of these applications are aided by intensity simulations using dynamical diffraction theory, henceforth referred to as dynamical simulations, which simulate multiple scattering using the Bloch wave approach and enable the fine details of a BKD pattern to be reproduced [33–36]. As with other electron diffraction modalities, studies have also utilized deep learning to aid EBSD analysis [37–47].

A drawback of these methods is dependence on prior knowledge of the sample phases [48], which may not be available after high-throughput synthesis of unknown samples, so they are not optimal for crystal symmetry determination. While there are methods to classify the crystal symmetry with manual analysis [49–54], they may be too time-consuming for analyzing a large volume of patterns. Furthermore, they do not give any space group predictions because of the complexity and orientation dependency of Kikuchi-band profiles, and there have not been any fully successful demonstrations of ab-initio space group classification in the literature. [53,55,56]. Neural network methods have been attempted for both space group and Bravais lattice classification, but difficulties tend to arise when evaluating models on phases outside the training set (SI Table 1)[57,58].

In this study, we present a framework for training neural networks for analyzing space groups. This is accomplished through a high-throughput dynamical BKD simulation dataset, paired with domain adaptation-based training, which enables the space group classification of experimental BKD patterns. The space group types 221 ($Pm\bar{3}m$), 223 ($Pm\bar{3}n$), 225 ($Fm\bar{3}m$), 227 ($Fd\bar{3}m$), 229 ($Im\bar{3}m$), and 230 ($Ia\bar{3}d$) from the cubic point group $m\bar{3}m$ are included in this study. First, simulated training data is obtained through dynamical simulations of crystal structures from the Materials Project [59] database. Next, models are trained simultaneously on simulated and experimental patterns to make predictions for experimental data through unsupervised domain adaptation. Unsupervised domain adaptation is a technique in deep learning which is used for simultaneous training on a source dataset, for which labels are provided, and a target dataset, for which labels are absent. While the target dataset lacks labels, it contains the same set of classes as the source dataset, such that a deep learning model could extract information from the labeled source dataset to train on the target dataset[60].

**Results**

Our workflow is outlined in Figure 1. First, crystal structures are queried from the Materials Project database and their BKD patterns are simulated using EMsoft [61]. Next, convolutional neural network (CNN) models are trained and tested on simulated patterns to analyze theoretical performances, and finally trained on a combination of experimental and simulated patterns through a deep learning technique known as unsupervised domain adaptation in order to classify experimental patterns.

When simulating the dataset, to optimize computational efficiency, crystal structures with the lowest lattice parameters from each space group type were selected. All crystal structures in the database with a unit cell volume less than 250 Å³ were simulated for space group types $Pm\bar{3}m$, $Pm\bar{3}n$, $Fm\bar{3}m$, and $Im\bar{3}m$. For $Fd\bar{3}m$, all structures with a unit cell volume less than 500 Å³ were selected, and for $Ia\bar{3}d$, all crystal structures with a volume less than 2000 Å³ were simulated, due to the naturally larger lattice parameters prevalent in these space group types.

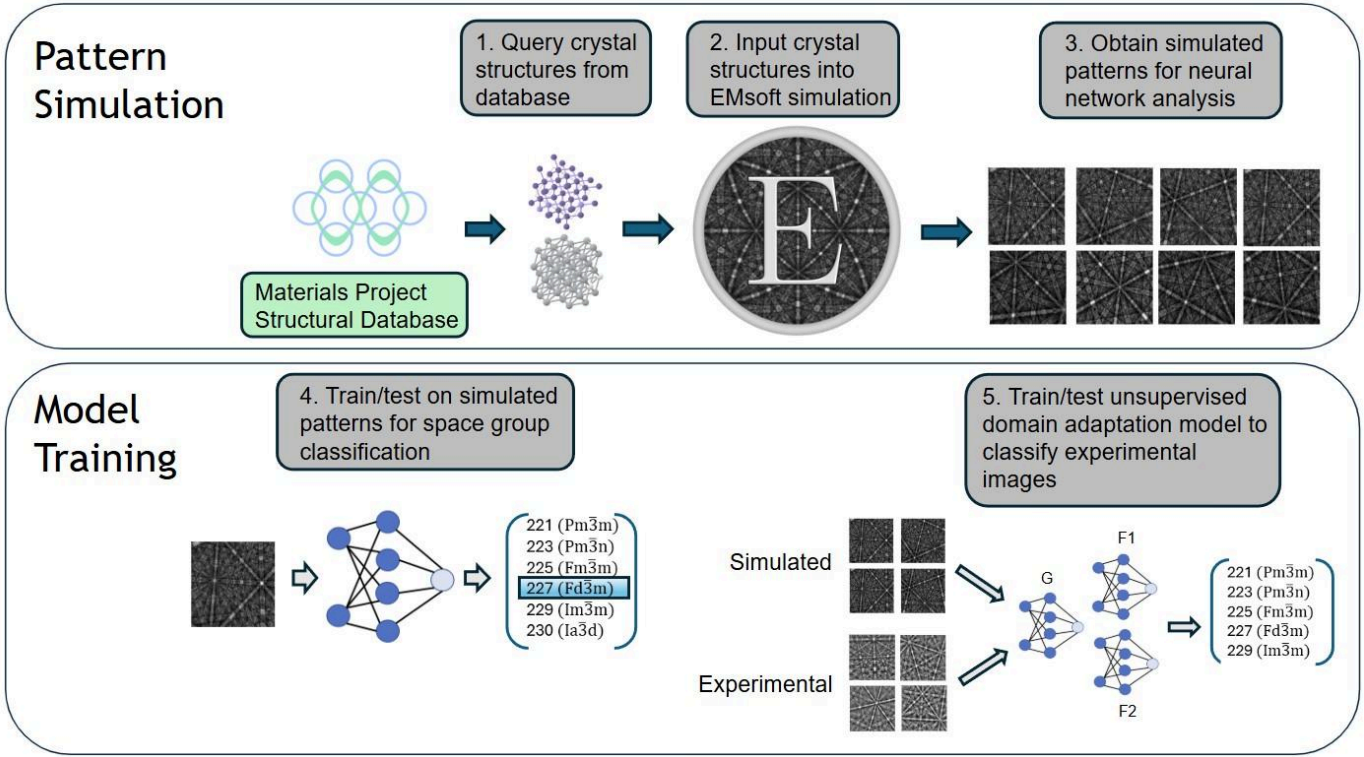

Figure 1: Workflow for developing space group type classification models

Next, when training and testing on simulated patterns, we show that in order to optimize neural network accuracy, it is necessary to relabel for each phase the space group obtained after compositional disordering, i.e., when all the atoms in the unit cell are set to the same atomic number. When models are classified to predict the true space group, as in previous studies[58], misclassifications tend to occur. This was seen when we trained and tested a convolutional neural network classification model with the Resnet18 architecture[62] on only simulated patterns to predict the true space group. Here, the models were evaluated using 5-fold cross validation, and each fold contained a distinct set of phases in order to prevent data leakage. An accuracy of 91% was first seen.

A confusion matrix, displayed in Figure 2a, is calculated from the patterns from all 5 validation splits and their corresponding predictions from the model trained on their respective training datasets. It is observed that $Pm\bar{3}m$, $Fm\bar{3}m$, and $Im\bar{3}m$ are most frequently misclassified for each other. These three types are frequently k-type maximal subgroups of each other [63] (SI Figure 2), which suggests that phases in these space group types may fluidly transition between the three space group types whenever it loses or gains lattice centering. This mostly happens from changes in compositional ordering. For example, in a B2 structure, if the constituent elements become compositionally disordered, the structure gains body-centered symmetry and the space group type transitions from Pm$\bar{3}$m to Im$\bar{3}$m.

Accordingly, if the compositional ordering is difficult to distinguish from a BKD pattern, then it will be difficult to classify the space group type correctly. This is expected to occur if there is a low difference in scattering factor between different elements in a phase. Previous works on lattice determination of BKD patterns [54] have indeed found that pseudosymmetry frequently occurs when the differences in atomic scattering factors or atomic numbers of different elements in a crystal structure are too small or too large. Our machine learning model displays similar trends. For example, for B2 structures, materials with lower differences in atomic number between the constituent elements are slightly more likely to be misclassified. This can be seen in Figure 2a), where misclassifications for different B2-type materials in the validation datasets are plotted as a function of $\Delta Z$ (the difference in atomic number between the two elements) and $\bar{Z}$ (the average atomic number of the two elements). Here, large clusters of misclassifications are seen at the bottom of the plot, where $\Delta Z$ is small. From Figure 2b), the deep learning models seem less susceptible to misclassifying B1 structures. B1 structures are expected to be misclassified as $A_h$ type structures (Pm-3m), since a B1 type structure transforms into an $A_h$ type structure after compositional disordering. However, the $A_h$ type structure occurs infrequently in the dataset (N=18), such that the model will still be biased to predict B1 type structures as having B1 symmetry.

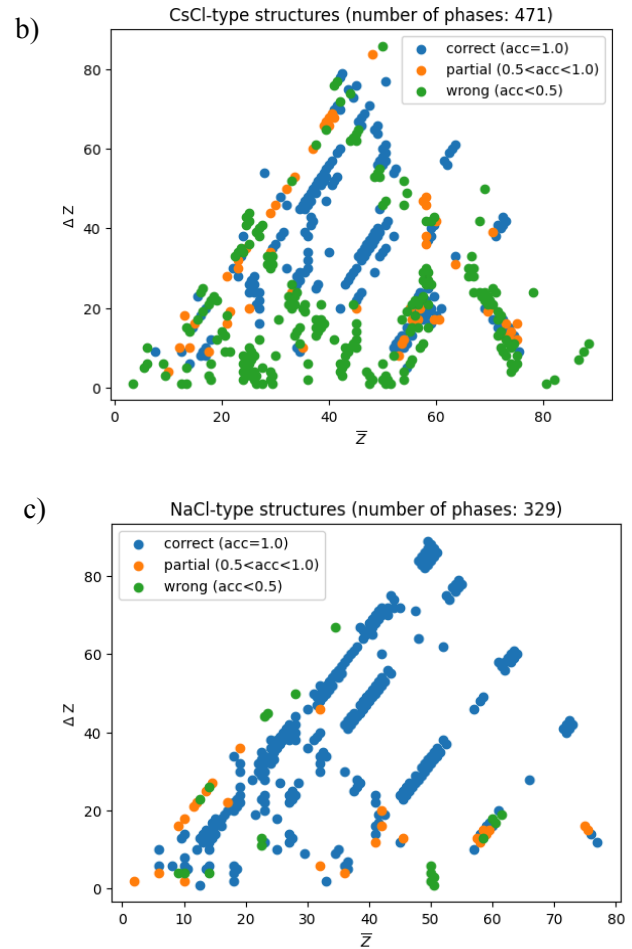

Figure 2: a) Normalized confusion matrix of model predictions during cross-validation. b) Plot of misclassifications for CsCl structured materials as a function of $\Delta Z$ and $\overline{Z}$. c) Plot of misclassifications for NaCl structured materials.

Next, through manual pattern analysis via the *CALM* software[50], we note that the patterns labeled Ia$\overline{3}$d in the experimental data, which was originally generated by Kaufmann et al.[58], often failed to show the appropriate I lattices. There are two phases in the space group type Ia$\overline{3}$d, which were Al4Ni3 and Al4CoNi2, which are superstructures based on a B2 structure prototype containing ordered vacancies with a periodicity 4 times the lattice parameter of the B2 sublattice. However, many patterns from Al4Ni3 fitted to a hexagonal lattice (SI Figure 3) with trigonal symmetry. This matches to phase Al3Ni2, so it is possible that this phase locally occurred in the sample. Furthermore, some Al4Ni3 patterns even contain visible bands that occur in simulated patterns of Al3Ni2 and are absent in Al4Ni3 simulations, for which an example is in SI Figure 4. Next, many patterns from Al4CoNi2 fitted to a P lattice (SI Figure 5), with a lattice parameter close to that of NiAl, such that superstructure reflections from ordered vacancies were not found in the pattern. However, in the patterns fitted to a P lattice, superstructure reflections caused by the chemical ordering of Al and (Co,Ni) were still found, such as in Si Figure 6, suggesting that the underlying B2 structure is still detectable. The difficulty of detecting superstructure reflections in BKD patterns, especially those that increase the lattice parameter relative to the original substructure, aligns with trends in the literature[54].

Overall, the error profile of deep learning and manual analysis have similarities, as they are both driven by low differentiability in the atomic scattering factors of different elements. Furthermore, there are difficulties detecting superstructure reflections that increase the lattice parameter. These constraints motivate our relabeling scheme for space group determination, where we train our model to predict the space group of the equivalent compositionally disordered structure, obtained when all the atoms in the unit cell are changed to the same element. Furthermore, we focus only on the space group types Pm$\overline{3}$m, Pm$\overline{3}$n, Fm$\overline{3}$m, Fd$\overline{3}$m, and Im$\overline{3}$m, since the superstructure reflections from Ia$\overline{3}$d are not reliably visible.

To evaluate this scheme, a Resnet18 model was trained on a simulated cross-validation dataset to predict the compositionally disordered space group type. The overall cross-validation accuracy is high at 98%, with a confusion matrix shown in Figure 3a). The remaining misclassified materials tend to contain very light elements, for which the scattering factor may be too small to contribute significantly to the pattern. This is expected because of the Z-contrast of backscattered electrons, where light elements have a lower backscatter yield compared to heavier elements. Out of the 95 crystal structures in the entire dataset for which the individual cross-validation accuracy was below 50%, 80 contained an element with an atomic number less than 10. The most common of these light elements were H (29 structures), Li (31 structures), and O (16 structures). Of the remaining misclassified structures, more common crystal structures were CaF2-type structures, for which there was an 81% overall cross-validation accuracy across the entire dataset. This is likely due to the presence of diamond cubic structures (Fd-3m) in the dataset, which are similar to CaF2-type structures–it can be seen in Figure 3b-3d that after a CaF2 structure is compositionally disordered, the only difference from an A4 structure is ordered vacancies in the 8c Wyckoff position. Finally, some structures with more unusual structure prototypes, like MnGa4 (mp-1069288); and some elemental structures like Se (mp-12771), K (mp-998881), Sr (mp-639774), U

(mp-1056699), Br (mp-1062055), and Ne (mp-111). There are only 19 elemental crystal structures in the space group $Pm\bar{3}m$, and 6 diamond cubic structures, and their scarcity may increase the probability of misclassification.

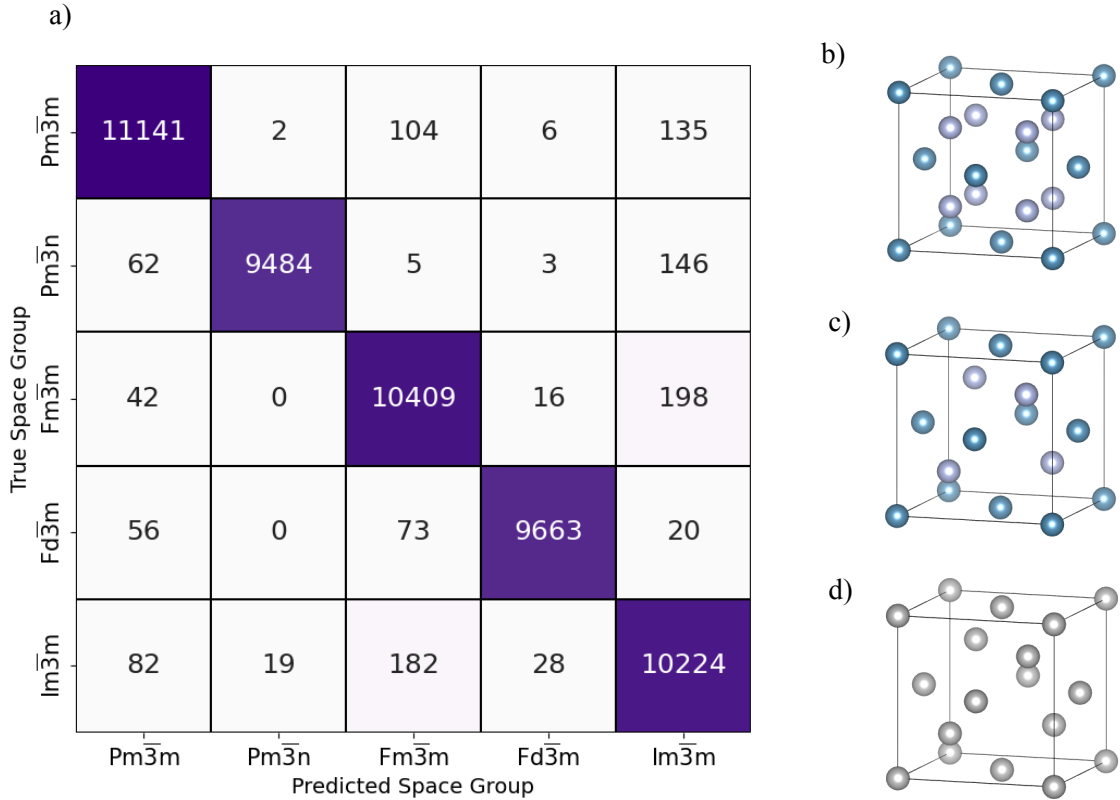

Figure 3: a) Confusion matrix displaying cross-validation Resnet18 accuracies on predicting the space group of the equivalent compositionally disordered structure. b) Unit cell of $CaF_2$ (fluorite). c) Unit cell of $CaF_2$ with vacancies in the 8c Wyckoff position, equivalent to a B3 structure. d) Unit cell for a diamond structure (A4 prototype) after compositional disordering.

However, the cross-validation score is still high and suggests that misclassifications are in theory very rare, at least for simulated patterns with very high band contrasts. It is an improvement over the 91% cross-validation score achieved when the model was trained to predict the true space group type, and the increase in accuracy confirms that differences in compositional ordering are a large cause of misclassifications when classifying the true crystal symmetry. When classifying the compositionally disordered space group, the final models show strong capabilities at differentiating between phases in the same Bravais lattice. In this scheme, the neural network is able to extract 3-dimensional information on an elemental structure prototype, and knowledge of compositional ordering acquired separately can help determine the final crystal structure.

Next, Maximum Classifier Discrepancy (MCD) [64], which is an unsupervised domain adaptation technique, is used to train a model to predict the space group type for experimental patterns, using both the labeled simulated data and the unlabeled experimental data for training. The experimental training and testing data used in this study is reported to contain BKD patterns from 18 different phases described by the 6 aforementioned space group types. A table showing the phases included, their true space groups, and their new labels obtained from compositional disordering is shown in Table 1. Al4Ni3 is excluded because its patterns are not reliably cubic, and Al4CoNi2 is included but relabeled to $Im\bar{3}m$ (bcc), which is in accordance with results from CALM analysis.

| Composition | True space group | New label | Composition | True space group | New label |
|---|---|---|---|---|---|
| FeNi$_3$ | $Pm\bar{3}m$ | $Fm\bar{3}m$ | TiC | $Fm\bar{3}m$ | $Pm\bar{3}m$ |
| Ni$_3$Al | $Pm\bar{3}m$ | $Fm\bar{3}m$ | Al | $Fm\bar{3}m$ | $Fm\bar{3}m$ |
| NiAl | $Pm\bar{3}m$ | $Im\bar{3}m$ | Ge | $Fd\bar{3}m$ | $Fd\bar{3}m$ |
| FeAl | $Pm\bar{3}m$ | $Im\bar{3}m$ | Si | $Fd\bar{3}m$ | $Fd\bar{3}m$ |
| Mo$_3$Si | $Pm\bar{3}n$ | $Pm\bar{3}n$ | W | $Im\bar{3}m$ | $Im\bar{3}m$ |
| Cr$_3$Si | $Pm\bar{3}n$ | $Pm\bar{3}n$ | Ta | $Im\bar{3}m$ | $Im\bar{3}m$ |
| TaC | $Fm\bar{3}m$ | $Pm\bar{3}m$ | Fe | $Im\bar{3}m$ | $Im\bar{3}m$ |
| Ni | $Fm\bar{3}m$ | $Fm\bar{3}m$ | Al$_4$CoNi$_2$ | $Ia\bar{3}d$ | $Im\bar{3}m$ |
| NbC | $Fm\bar{3}m$ | $Pm\bar{3}m$ | Al$_4$Ni$_3$ | $Ia\bar{3}d$ | – |

Table 1: Table of phases in the experimental data, with their space groups and new recalculated space groups after compositional disordering

First, MCD is used to train neural networks on patterns from all the phases, and then evaluated on new patterns from the same phases, in order to evaluate its ability to match the unlabeled training patterns to their correct labels. The MCD models are trained accordingly for 30 different runs for statistical robustness, and the averages are reported at a 95% confidence interval. The average test accuracies are shown in Figure 4a, and the accuracies for individual materials are displayed in SI Figure 7a. Due to class imbalance[65] and small dataset sizes, there is variance in performance in the different runs. However, ensemble voting can increase the final accuracy and resolve the class imbalance issue by considering for each test image the most frequent prediction among the 30 models as a final overall prediction. This can help achieve an even higher final score.

Here, a correlation between IQ and average accuracy is observed–NiAl and Al have the lowest accuracies, which can be seen in Figure 4b. Furthermore, when their patterns are removed, the average accuracy increases significantly for the remaining phases (SI Figure 7). For NiAl and Al, it is likely that the low pattern quality is causing the misclassifications, because when a Resnet18 model is trained and tested on simulated patterns, NiAl and Al are correctly classified (SI Figure 8a). Furthermore, when only

Al is removed, and Al4CoNi2 is replaced with NiAl, the accuracy on NiAl is still close to zero, which is far less than that of Al4CoNi2 (SI Figure 9). From the previous CALM analysis, Al4CoNi2 and NiAl are expected to have virtually identical reciprocal lattices and Z contrasts, while the NiAl patterns have significantly lower pattern qualities, which further suggests that low pattern quality in NiAl is the reason for its low accuracy.

a)

|  | Average accuracy per material | Ensemble voting accuracy per material |
| --- | --- | --- |
| Including all phases | 0.71±0.01 | 0.73 |
| Removing low pattern quality phases (NiAl, Al) | 0.89±0.03 | 0.93 |

b)

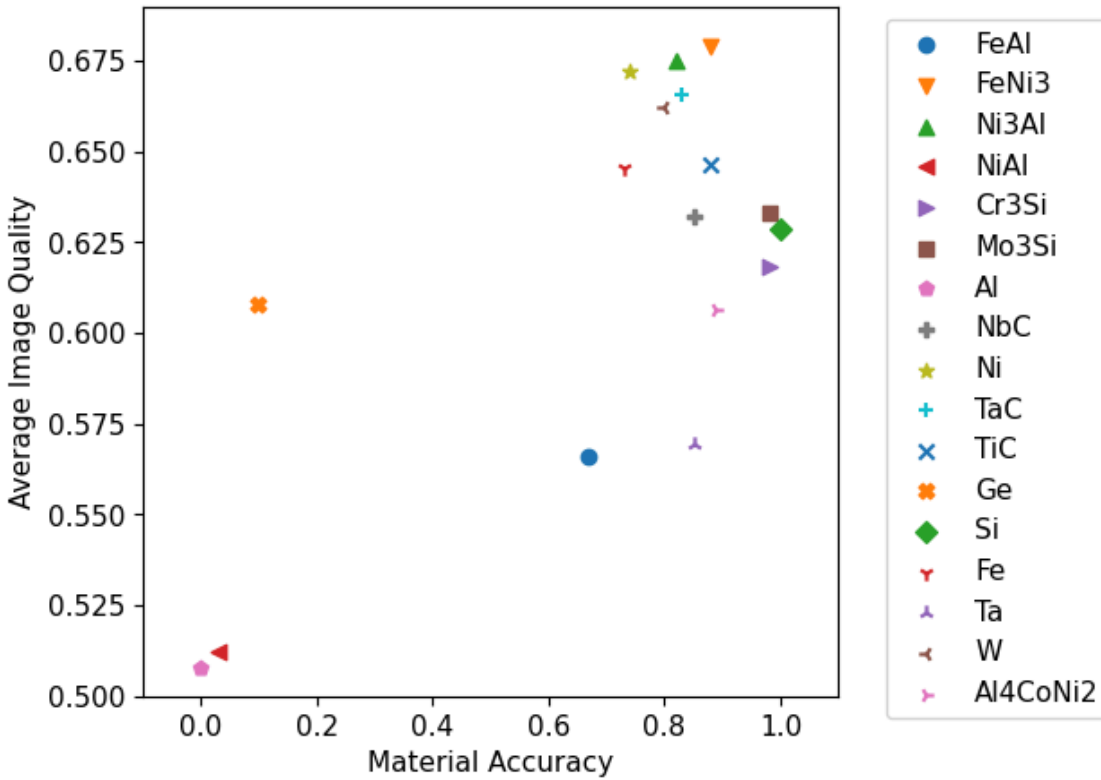

Figure 4a) MCD performance in 30 different runs when classifying the space group of the compositionally disordered equivalent structure. b) Scatter plot revealing the trend between average accuracy and Image Quality (IQ) when classifying the space group type resulting from compositional disordering.

Removing phases with low pattern qualities can optimize the accuracy. For example, Ge was previously misclassified before low-quality patterns were removed, and its accuracy greatly increased after the low-quality patterns were removed (SI Figure 7). Ge is also diamond-structured, and as noted previously, this makes it susceptible to being confused with fluorite-type structures. In fact, after training and testing a CNN on simulated data, the space group type $Fm\bar{3}m$ is incorrectly predicted for Ge and Si, but $Fd\bar{3}m$ is correctly predicted when the CaF2-type structures are removed (SI Figure 8). However, when training the MCD model, even when the CaF2 type structures are removed from the simulated training data, performance doesn't improve on the experimental Ge patterns until the NiAl and Al patterns are also removed (SI Figure 10).

**SUMMARY AND OUTLOOK**

Rapid crystal symmetry classification methods are needed for analyzing a large number of unknown phases in high-throughput crystalline nanomaterial discovery workflows. Paired with elemental compositions, the derived crystal symmetry information can enable the determination of crystal structures and enhance understanding of nanomaterial structure-property relationships. In this study, we presented a deep learning approach for classifying the space group type of BKD patterns, motivated by their prevalence of 3-dimensional crystal structure information that may not be accessible in other electron diffraction modalities, as well as the large sample chamber of the SEM which can hold centimeter-scale megalibrary chips.

In future studies, the insights from this study will be used to develop models that can be applied to a wider variety of symmetries, including extending models to other crystal systems. Strategies for anticipating and adapting to misclassifications based on crystallographic principles will also be applied. For example, the findings from this study suggest that training a model to predict the symmetry obtained from compositional disordering yielded a higher accuracy, as well as keeping unit cell volumes low due to the weak signal of superstructure reflections. When implementing the deep learning models, experimental work will also be done to optimize experimental pattern qualities and spatial resolutions.

Finally, it should be noted that the models here are trained and tested on the same phases–an upcoming version will present results showing the model performance on novel phases outside the training dataset.

**METHODS**

*High-throughput simulation*

EMsoft was used to generate dynamical simulations for the selected materials. First, .xtal files were created by reading data from each selected CIF file from the Materials Project database, which was automated using the pexpect python module. For the space group type $Fd\bar{3}m$, EMsoft requires the origin choice (1 or 2) to be specified. Thus, FINDSYM[66] was used to create a new CIF file for each $Fd\bar{3}m$ crystal structure, which had the origin choice explicitly specified in the file. Next, the .xtal files were used to generate a master pattern for each material. Dynamical master pattern simulations were attempted for a total of 5148 crystal structures. A small fraction of the materials in the space group type $Ia\bar{3}d$ that did not finish after a 12 hour job were left out of the dataset. Finally, gnomonic projections were simulated using the master patterns in order to create images for training, validation, and testing. The sample-detector geometry used for the projections was set to be as close as possible to that of the experimental data, which is described in Kaufmann et al.[67]. The sample-detector distance was set to 19100 microns, and the detector tilt was set to 13.7 degrees. The x and y coordinates of the pattern center and the pixel size were unknown. Thus, an arbitrary value for the pattern center from the literature was used [68], which was 171 microns above the center of the detector screen, and the pixel size was guessed to be 23 microns, which by inspection seemed to provide good agreement with experimental data. While close agreement between simulated and experimental parameters was desired, a perfect match was not required, as it was expected that the domain adaptation training would enable the neural networks to adjust to slight discrepancies.

Other described acquisition parameters used to collect the experimental data were replicated in the simulation. As the experimental accelerating voltage was 20 keV, each integer energy level from 10-20 keV was simulated. The simulation dwell time was set to 800 microseconds, and the beam current was set to 51 nA. The maximum depth was set to 100 nm, with a step size of 1 nm, which were the default values. To ensure a sufficient number of bands was included, the minimum d-spacing was set to 0.05 nm, and for the Bethe parameters, the strong beam cutoff used was 8.0, the weak beam cutoff was 50.0, and the complete cutoff was 100.0. Finally, to reduce simulation time, the dimensions of the square master pattern were lowered to 601x601 pixels (*npx*=300).

Some experimental patterns were found to have degradations in pattern quality, particularly those from Ta. The experimental patterns visibly lacked dynamical diffraction effects, as shown in SI Figure 11, and were more similar to simulated patterns with a higher B-factor (referred to as "Debye-Waller factor" in EMsoft). Although the B-factor is used to consider possible thermal vibrations, high apparent B-factors in experimental patterns are also known to result from deviations from an ideal theoretical structure. Thus, each simulated material in the space group types $Pm\bar{3}m$, $Pm\bar{3}n$, $Fm\bar{3}m$, $Fd\bar{3}m$, and $Im\bar{3}m$ was re-simulated, with each atom having a B-factor randomly selected from 0.005 to 0.011 $nm^2$, which were to be added to the training data.

Simulations were performed on a supercomputing cluster so that many could be performed in parallel. Furthermore, to decrease computation time, GPUs were used for the Monte Carlo simulations and 128 OpenMP threads were used for each master pattern calculation.

*Preparation of Simulated Datasets from Master Patterns*

Three datasets were prepared from the simulated master patterns: a simulated cross-validation dataset, a combined training dataset derived from combining all the images in the cross-validation dataset, as well as a simulated test dataset.

For the simulated patterns of the materials present in the experimental dataset, 100 gnomonic projections of random orientations were calculated from each master pattern to create a simulated test dataset. The master patterns of the rest of the materials were then used to generate a 5-fold cross-validation dataset: for each space group type, all the corresponding master patterns were randomly split into 5 folds, from which gnomonic projections of random orientations are calculated to create the 5 different train/validation datasets for 5-fold cross-validation. This ensured that none of the phases in a validation dataset would also be present in the corresponding training dataset, enabling the evaluation of model performance on novel phases. Furthermore, due to the imbalance in the number of crystal structures simulated for each space group type, each type had a different number of gnomonic projections sampled per master pattern, such that each space group type would have a similar total number of patterns.

After creating the test and cross-validation dataset, the final training dataset was created by combining the training and validation data from all the folds of the cross-validation dataset.

A training dataset from the master patterns that had varying B-factors was created in a similar way, to use during MCD training. Cross-validation and test datasets were not created from these master patterns.

*Resnet18 training and testing on simulated data*

Neural networks with the Resnet18 architecture pre-trained on ImageNet were first trained and tested on the simulated images with low B-factors. During training, the simulated images were resized to an image resolution of 229x229 pixels to be input into the model, which were randomly cropped into size 224x224 images during training. Data augmentation was also used by randomly flipping images up-down and left-right. Stochastic gradient descent was used for training, with a learning rate of 0.001 and with a momentum of 0.9. The learning rate decreased by 10% every 7 epochs. First, the model was trained on the cross-validation datasets for 10 epochs, after which the optimal number of epochs to train was determined from the epoch with the highest accuracy score. Then, a new model was trained on the respective combined training data and tested on the simulated test dataset. The optimal number of epochs was 10 for compositionally disordered space group type classification.

*MCD model architecture and training*

Maximum Classifier Discrepancy (MCD) was used to train a model with a ResNet50 backbone pre-trained on ImageNet[69] to predict the space group type of experimental BKD patterns using unsupervised domain adaptation, of which the implementation by Zhou et al. was used [70,71]. This method involves adversarial training between two discriminators and a generator. One training step contains three distinct stages: in one stage, target images are input into the generator, and the generator output is input into both the discriminators. With the weights of the generator frozen, the weights of the discriminators are then updated to maximize the discrepancy between the outputs of the two discriminators, as well as classify the source images by minimizing cross-entropy loss. The discrepancy is a measure of the dissimilarity between the two output vectors, and is calculated as the L1 distance between the pair of vectors.

In another stage, the discriminator weights are frozen and the generator is trained to minimize the discrepancy. This stage is repeated 4 times for each mini-batch (*n*=4), which was found to yield the highest performance in the original study. The final stage involves training both discriminator-generator pairs to classify the source images by minimizing cross-entropy loss. More specific details on the training method can be found in the original study[64].

In our implementation, the generator consists of a ResNet50 backbone, and the classifiers each consist of a single perceptron layer. Models were trained using the Adam optimizer with a learning rate of 0.0002 and batch size of 128, following the original implementation. The model was trained for 20 epochs. Additionally, no class balance loss term was used as done in the original MCD study, since the assumption of balanced classes in the experimental data did not hold.

During training, the simulated dataset consisted of both patterns with high B-factors, as well as patterns with low B-factors.

The experimental training dataset was created by combining the training and validation sets from Kaufmann et al.[58] A total of 144,465 patterns are in the testing dataset, and 3,000 patterns are contained in the combined training and validation datasets. To improve pattern quality, background correction and 30 pattern averaging was used during pattern acquisition in the original study. However, to ensure high pattern quality, the Image Quality (IQ)[72] of each pattern was calculated, and the 100 highest quality images were sampled from each phase for training.

*CALM analysis*

CALM was used to analyze the experimental patterns in the space group type $Ia\bar{3}d$ dataset. As many lattice plane traces and zone axes as possible are selected in each pattern, and then automatic bandwidth detection is applied to derive a lattice solution. However, because automated bandwidth selection could overlook local extrema in band profiles that represent lower-order interferences [54], lower order interferences were searched for afterwards to obtain another refined lattice solution which may be different from that obtained through automated bandwidth detection. An example of a band profile from an Al4CoNi2 pattern containing lower-order interferences is shown in SI Figure 6.

**Acknowledgements:**

This research was supported in part through the computational resources and staff contributions provided for the Quest high performance computing facility at Northwestern University which is jointly supported by the Office of the Provost, the Office for Research, and Northwestern University Information Technology. This work made use of Northwestern University's NUANCE Center, which has received support from the SHyNE Resource (NSF ECCS-2025633), the IIN, and Northwestern's MRSEC program (NSF DMR-2308691). The authors also acknowledge funding from the International Institute for Nanotechnology at Northwestern University, and the Predictive Science and Engineering Design cluster program supported by the Northwestern Interdisciplinary Cluster Initiative. The authors also thank Dr. Chad Mirkin, Dr. Wei Chen, and Dr. Daniel Apley for helpful discussions.

**Supplementary Material**

**Analysis of results from literature**

Kaufmann et al. trained neural network models to predict the Bravais lattice of materials from EBSD patterns [1]. In particular, Kaufmann et al. test their models on patterns from 9 different samples whose patterns were not included in their training dataset. As shown in Table 1, their model predicts silicon patterns as cP, and Mo2C patterns as mP, which were counted as correct classifications. However, silicon is mostly known in the literature to be diamond cubic, and Mo2C is commonly reported to be oP [2], and structures with the reported lattices were not found in the Materials Project database. Therefore, it is unclear whether these materials are being classified correctly by the model.

| Material | Reported accuracy | Most commonly predicted lattice | Reported ground truth lattice | Lattices of Materials Project structures (synthesized structures only) |
|---|---|---|---|---|
| Si | 83.7% | cP | cP | hP (mp-165, mp-34, mp-10649), oC (mp-1095269, mp-1079649), cI (mp-168), hR (mp-571520), tI (mp-92, mp-1056579), cF (mp-149, mp-16220, mp-27). |
| $Mo_2C$ | 92.6% | mP | mP | oP (mp-1552) |

**Table 1**: Discrepancies in ground truth between online structural database (Materials Project) and reported Bravais lattice

| Group | Maximal subgroups (index, type) |
|---|---|
| Pm-3m | Fm-3m (2,k); Im-3m (4,k) |
| Pm-3n | Ia-3d (4,k) |
| Fm-3m | Pm-3m (4, k) |
| Fd-3m | – |
| Im-3m | Pm-3m (2, k); Pm-3n (2,k) |
| Ia-3d | – |

**Figure 2**: Table of maximal subgroups as well as their respective indices. Only the 6 space group types in this study are included. These space group types are *klassengleichen* subgroups of each other [3], [4], [5].

**CALM analysis**

| pattern name | used bands | a | b | c | α | β | γ | ϕ | δ |
|---|---|---|---|---|---|---|---|---|---|
| 2637 | 50 | 2.867 | 2.867 | 2.872 | 90.1 | 90 | 89.8 | 0.1 | 0.001 |
| 2740 | 55 | 3.998 | 3.995 | 4.871 | 90 | 90 | 119.8 | 0.07 | 0.001 |
| 2774 | 51 | 4.037 | 4.029 | 4.92 | 90.2 | 89.9 | 119.8 | 0.16 | 0.002 |
| 2937 | 59 | 4.02 | 4.011 | 4.903 | 90 | 90.1 | 120 | 0.03 | 0.002 |
| 2942 | 87 | 4.021 | 4.02 | 4.882 | 90.1 | 89.9 | 119.9 | 0.08 | 0 |
| 3027 | 67 | 4.024 | 4.019 | 4.914 | 90 | 90 | 119.9 | 0.06 | 0.001 |
| 3051 | 46 | 4.019 | 4.018 | 4.862 | 90 | 89.9 | 119.9 | 0.07 | 0 |
| 3054 | 25 | 2.861 | 2.861 | 2.864 | 90 | 98.9 | 89.9 | 0.05 | 0.001 |
| 3060 | 70 | 4.001 | 3.984 | 4.863 | 90 | 90.2 | 120 | 0.07 | 0.004 |
| 3063 | 62 | 4.006 | 3.999 | 4.871 | 90.2 | 90 | 120 | 0.07 | 0.002 |
| 3103 | 55 | 4.019 | 4.012 | 4.895 | 90.3 | 89.9 | 120 | 0.13 | 0.002 |
| 3106 | 72 | 4.019 | 4.014 | 4.875 | 90.1 | 89.9 | 120 | 0.09 | 0.001 |
| 3109 | 68 | 3.998 | 3.991 | 4.869 | 90 | 89.9 | 119.9 | 0.07 | 0.002 |
| 3115 | 53 | 4.017 | 4.016 | 4.876 | 90.2 | 89.9 | 119.9 | 0.16 | 0 |
| 3162 | 50 | 4.022 | 4.007 | 4.896 | 90.2 | 89.9 | 119.9 | 0.12 | 0.004 |
| 3176 | 42 | 4.024 | 4.007 | 4.902 | 89.9 | 90.1 | 120 | 0.06 | 0.004 |
| 3181 | 56 | 4.01 | 4.006 | 4.889 | 90 | 90.1 | 120 | 0.03 | 0.001 |
| 3240 | 63 | 4.042 | 4.036 | 4.93 | 90 | 90 | 120 | 0.02 | 0.001 |

**Figure 3**: Results of CALM analysis for a handful of Al4Ni3 patterns chosen randomly. The lattice parameters a, b, and c are given as well as the angles α, β, and γ. Φ describes the deviation in lattice parameter ratio from the ideal lattice, and δ describes the deviation in lattice angles from the ideal values.

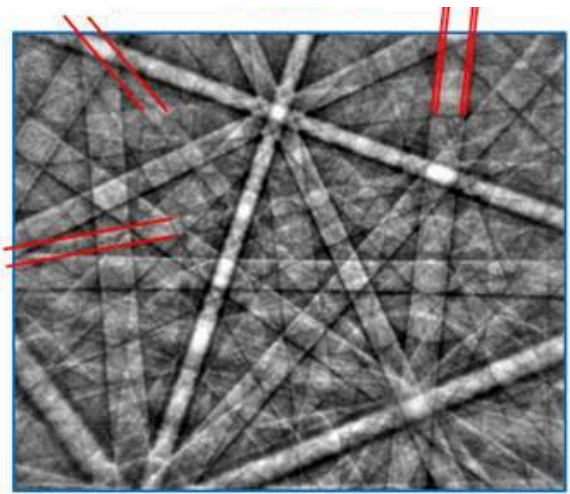
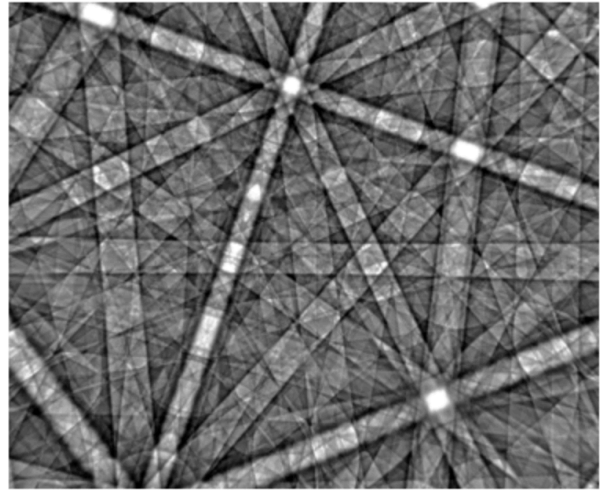
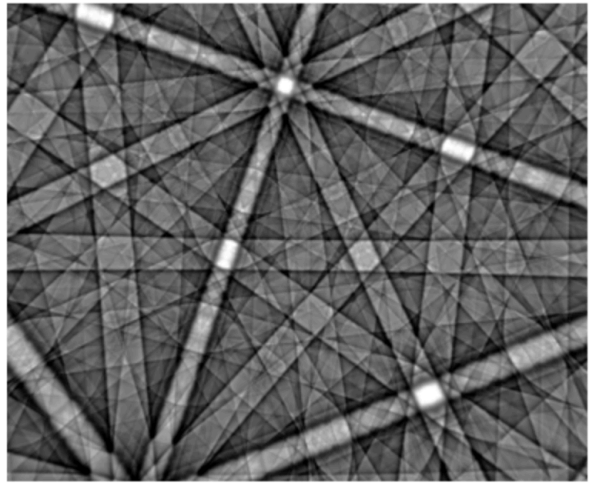

**Figure 4:** An experimental pattern displayed with corresponding DynamicS (Bruker) simulations of $Al_3Ni_2$ and $Al_4Ni_3$, along with the normalized cross-correlation score. The $Al_3Ni_2$ has a higher score, suggesting greater similarity with the experimental pattern. Some visible features in the experimental pattern that can be seen exclusively in the $Al_3Ni_2$ simulation are highlighted in red.

| pattern name | used bands | a | b | c | α | β | γ | ϕ | δ | Lattice type |
|---|---|---|---|---|---|---|---|---|---|---|
| 0693 | 59 | 2.835 | 2.841 | 2.842 | 90 | 90 | 90 | 0.02 | 0.001 | cP |
| 0704 | 41 | 2.837 | 2.839 | 2.839 | 90 | 90 | 90.1 | 0.04 | 0 | cP |
| 0744 | 46 | 2.837 | 2.842 | 2.844 | 90 | 90 | 90 | 0.03 | 0.001 | cP |
| 0749 | 49 | 2.813 | 2.814 | 2.816 | 90.1 | 89.9 | 89.9 | 0.10 | 0 | cP |
| 0756 | 51 | 2.847 | 2.848 | 2.849 | 90 | 90 | 90.1 | 0.06 | 0 | cP |
| 0762 | 35 | 2.85 | 2.845 | 2.853 | 89.8 | 90.0 | 90.1 | 0.11 | 0.001 | cI |
| 0766 | 42 | 5.347 | 5.589 | 4.916 | 73.1 | 89.8 | 118.4 | n/a | n/a | aP |
| 0769 | 34 | 2.84 | 2.842 | 2.848 | 90 | 90 | 89.9 | 0.05 | 0.001 | cP |
| 0824 | 34 | 2.826 | 2.827 | 2.832 | 89.7 | 90 | 89.9 | 0.12 | 0.001 | cP |
| 0994 | 38 | 8.038 | 8.042 | 5.686 | 90 | 89.9 | 90.1 | 0.06 | 0.001 | tP |
| 1490 | 49 | 2.837 | 2.838 | 2.839 | 90 | 90 | 89.9 | 0.05 | 0 | cP |
| 1600 | 43 | 2.842 | 2.846 | 2.846 | 89.9 | 90.1 | 89.9 | 0.09 | 0.001 | cP |
| 0878 | 41 | 2.815 | 2.821 | 2.823 | 90 | 90 | 90.1 | 0.03 | 0.003 | cI |
| 1264 | 32 | 2.822 | 2.828 | 2.828 | 90 | 89.8 | 90 | 0.08 | 0.001 | cI |

**Figure 5**: Lattice solutions of Al4CoNi2 patterns, along with lattice parameters and angles, as well as deviations ϕ and δ from ideal lattice parameter ratios and ideal lattice angles.

## CALM analysis of patterns from Al4CoNi2

a)

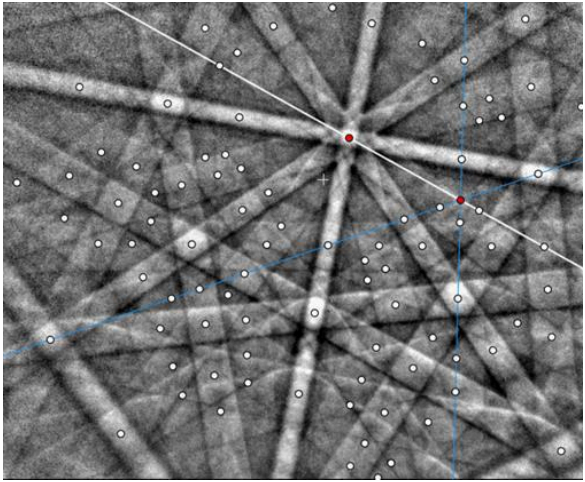

b)

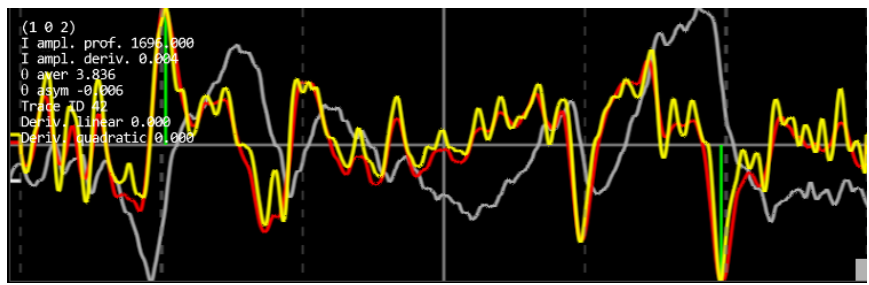

**Figure 6:** a) One example of an EBSD pattern from the Al4CoNi2 dataset. Selected zone axes are marked by the white dots. The band profile of (102) (white line trace) is shown in b), where the global extrema are marked by green lines, but a prominent lower order reflection is visible but not global.

a)

|         | Average score |
|---------|---------------|
| NiAl    | 0.02±0.01     |
| FeAl    | 0.7±0.04      |
| FeNi3   | 0.82±0.06     |
| Ni3Al   | 0.79±0.04     |
| Cr3Si   | 0.97±0.03     |
| Mo3Si   | 0.98±0.01     |
| Al      | 0.0±0.0       |
| NbC     | 0.83±0.02     |
| Ni      | 0.64±0.07     |
| TaC     | 0.78±0.04     |
| TiC     | 0.89±0.02     |
| Ge      | 0.09±0.02     |
| Si      | 1.0±0.0       |
| Fe      | 0.75±0.02     |
| Ta      | 0.85±0.02     |
| W       | 0.75±0.07     |
| Al4CoNi2 | 0.89±0.1     |

b)

|         | Average score |
|---------|---------------|
| FeAl    | 0.98±0.01     |
| FeNi3   | 0.88±0.04     |
| Ni3Al   | 0.87±0.03     |
| Cr3Si   | 0.97±0.01     |
| Mo3Si   | 0.96±0.02     |
| NbC     | 0.83±0.02     |
| Ni      | 0.8±0.05      |
| TaC     | 0.74±0.03     |
| TiC | 0.9±0.01 |
| Ge | 0.94±0.03 |
| Si | 0.99±0.0 |
| Fe | 0.89±0.03 |
| Ta | 0.98±0.01 |
| W | 0.92±0.09 |
| Al4CoNi2 | 0.84±0.15 |

**Figure 7**: Average accuracy breakdown of MCD on each phase in the experimental test data over 20 runs, with a 95% confidence interval given. a) With all phases. b) With patterns from NiAl and Al removed from experimental data.

a)

|   | Phase | Accuracy | Ground truth (with chemical disordering) | Most frequent prediction |
|---|---|---|---|---|
| 0 | FeNi3 | 1 | Fm-3m | Fm-3m |
| 1 | FeAl | 1 | Im-3m | Im-3m |
| 2 | NiAl | 1 | Im-3m | Im-3m |
| 3 | Ni3Al | 1 | Fm-3m | Fm-3m |
| 4 | Mo3Si | 1 | Pm-3n | Pm-3n |
| 5 | Cr3Si | 1 | Pm-3n | Pm-3n |
| 6 | NbC | 1 | Pm-3m | Pm-3m |
| 7 | TiC | 1 | Pm-3m | Pm-3m |
| 8 | TaC | 1 | Pm-3m | Pm-3m |
| 9 | Al | 1 | Fm-3m | Fm-3m |
| 10 | Ni | 1 | Fm-3m | Fm-3m |
| 11 | Ge | 0.24 | Fd-3m | Fm-3m |
| 12 | Si | 0.01 | Fd-3m | Fm-3m |
| 13 | W | 1 | Im-3m | Im-3m |
| 14 | Ta | 1 | Im-3m | Im-3m |
| 15 | Fe | 1 | Im-3m | Im-3m |
| Average | - | 0.89 | - | - |

b)

| | Phase | Accuracy | Ground truth (with chemical disordering) | Most frequent prediction |
|---|---|---|---|---|
| 0 | FeNi3 | 1 | Fm-3m | Fm-3m |
| 1 | FeAl | 1 | Im-3m | Im-3m |
| 2 | NiAl | 1 | Im-3m | Im-3m |
| 3 | Ni3Al | 1 | Fm-3m | Fm-3m |
| 4 | Mo3Si | 1 | Pm-3n | Pm-3n |
| 5 | Cr3Si | 1 | Pm-3n | Pm-3n |
| 6 | NbC | 1 | Pm-3m | Pm-3m |
| 7 | TiC | 1 | Pm-3m | Pm-3m |
| 8 | TaC | 1 | Pm-3m | Pm-3m |
| 9 | Al | 1 | Fm-3m | Fm-3m |
| 10 | Ni | 1 | Fm-3m | Fm-3m |
| 11 | Ge | 1 | Fd-3m | Fd-3m |
| 12 | Si | 0.89 | Fd-3m | Fd-3m |

| 13 | W | 1 | Im-3m | Im-3m |
| 14 | Ta | 1 | Im-3m | Im-3m |
| 15 | Fe | 1 | Im-3m | Im-3m |
| Average | - | 0.99 | - | - |

**Figure 8:** Performance of Resnet18 on predicting the space group type of the chemically disordered equivalent structure when the model is tested on simulated patterns of materials in the experimental dataset. a) With fluorite-type structures in training data b) Without fluorite-type structures in training data

|       | 0    | 1    | 2    | 3    | 4    | 5    | 6    | mean |
|-------|------|------|------|------|------|------|------|------|
| NiAl  | 0.06 | 0.02 | 0.06 | 0.01 | 0.04 | 0.02 | 0.02 | 0.03 |
| FeAl  | 0.84 | 0.7  | 0.85 | 0.75 | 0.87 | 0.71 | 0.71 | 0.78 |
| FeNi3 | 0.9  | 0.91 | 0.95 | 0.78 | 0.89 | 0.76 | 0.62 | 0.83 |
| Ni3Al | 0.85 | 0.83 | 0.89 | 0.73 | 0.84 | 0.76 | 0.68 | 0.8  |
| Cr3Si | 1    | 0.91 | 0.94 | 0.98 | 1    | 0.99 | 0.83 | 0.95 |
| Mo3Si | 1    | 0.91 | 0.93 | 0.96 | 0.99 | 0.98 | 0.91 | 0.95 |
| NbC   | 0.85 | 0.83 | 0.85 | 0.79 | 0.89 | 0.85 | 0.86 | 0.85 |
| Ni    | 0.72 | 0.73 | 0.82 | 0.59 | 0.8  | 0.5  | 0.48 | 0.66 |
| TaC   | 0.8  | 0.8  | 0.8  | 0.77 | 0.82 | 0.79 | 0.84 | 0.8  |
| TiC   | 0.92 | 0.9  | 0.92 | 0.8  | 0.92 | 0.89 | 0.9  | 0.89 |
| Ge    | 0.06 | 0.08 | 0.26 | 0.18 | 0.93 | 0.07 | 0.08 | 0.24 |
| Si    | 0.97 | 1    | 1    | 1    | 0.99 | 1    | 1    | 0.99 |
| Fe    | 0.75 | 0.75 | 0.79 | 0.73 | 0.89 | 0.73 | 0.67 | 0.76 |
| Ta    | 0.93 | 0.86 | 0.91 | 0.83 | 0.89 | 0.83 | 0.84 | 0.87 |
| W     | 0.85 | 0.81 | 0.9  | 0.69 | 0.95 | 0.72 | 0.66 | 0.8  |

**Figure 9**: Performance of MCD for 7 runs on predicting the space group of the chemically disordered equivalent structure when Al4CoNi2 and Al are removed from the experimental data, and NiAl is included

a)

|       | 0    | 1    | 2    | 3    | 4    | 5    | 6    | 7    | mean |
|-------|------|------|------|------|------|------|------|------|------|
| NiAl  | 0.02 | 0.09 | 0.02 | 0.04 | 0.01 | 0    | 0.02 | 0    | 0.02 |
| FeAl  | 0.62 | 0.58 | 0.78 | 0.45 | 0.62 | 0.4  | 0.53 | 0.74 | 0.59 |
| FeNi3 | 0.55 | 0.94 | 0.79 | 0.89 | 0.68 | 0.58 | 0.74 | 0.93 | 0.76 |
| Ni3Al | 0.59 | 0.86 | 0.79 | 0.85 | 0.61 | 0.68 | 0.75 | 0.86 | 0.75 |
| Cr3Si | 1    | 0.98 | 0.97 | 1    | 1    | 0.97 | 1    | 0.99 | 0.99 |
| Mo3Si | 1    | 0.97 | 0.95 | 0.97 | 1    | 0.97 | 1    | 0.98 | 0.98 |
| Al    | 0    | 0    | 0    | 0    | 0    | 0    | 0    | 0    | 0    |

|   | 0 | 1 | 2 | 3 | 4 | 5 | 6 | 7 | mean |
|---|---|---|---|---|---|---|---|---|---|
| NbC | 0.82 | 0.81 | 0.89 | 0.85 | 0.87 | 0.76 | 0.83 | 0.83 | 0.83 |
| Ni | 0.41 | 0.74 | 0.72 | 0.67 | 0.52 | 0.31 | 0.5 | 0.78 | 0.58 |
| TaC | 0.8 | 0.61 | 0.87 | 0.86 | 0.83 | 0.73 | 0.79 | 0.74 | 0.78 |
| TiC | 0.77 | 0.91 | 0.93 | 0.93 | 0.9 | 0.63 | 0.9 | 0.92 | 0.86 |
| Ge | 0.14 | 0.14 | 0.07 | 0.06 | 0.04 | 0.23 | 0.26 | 0.19 | 0.14 |
| Si | 1 | 1 | 1 | 0.99 | 1 | 1 | 1 | 1 | 1 |
| Fe | 0.74 | 0.69 | 0.78 | 0.65 | 0.68 | 0.57 | 0.7 | 0.74 | 0.69 |
| Ta | 0.89 | 0.79 | 0.85 | 0.59 | 0.9 | 0.79 | 0.82 | 0.84 | 0.81 |
| W | 0.87 | 0.59 | 0.86 | 0.47 | 0.88 | 0.59 | 0.78 | 0.82 | 0.73 |
| Al4CoNi2 | 0.91 | 0.95 | 0.98 | 0.96 | 0.9 | 0.86 | 0.94 | 0.96 | 0.93 |

b)

|   | 0 | 1 | 2 | 3 | 4 | 5 | 6 | 7 | mean |
|---|---|---|---|---|---|---|---|---|---|
| FeAl | 0.99 | 0.99 | 0.96 | 0.01 | 0.99 | 0.98 | 0 | 1 | 0.74 |
| FeNi3 | 0.91 | 0.93 | 0.46 | 0.53 | 0.66 | 0.97 | 0.12 | 0.64 | 0.65 |
| Ni3Al | 0.9 | 0.92 | 0.55 | 0.75 | 0.69 | 0.93 | 0.23 | 0.6 | 0.7 |
| Cr3Si | 0.99 | 0.95 | 0.94 | 0.98 | 0.99 | 0.99 | 0.99 | 0.97 | 0.98 |
| Mo3Si | 0.99 | 0.93 | 0.93 | 0.99 | 0.99 | 0.98 | 0.98 | 0.95 | 0.97 |
| NbC | 0.81 | 0.79 | 0.9 | 0.85 | 0.81 | 0.84 | 0.7 | 0.81 | 0.81 |
| Ni | 0.84 | 0.88 | 0.32 | 0.44 | 0.64 | 0.89 | 0.1 | 0.6 | 0.59 |
| TaC | 0.72 | 0.64 | 0.8 | 0.82 | 0.69 | 0.75 | 0.62 | 0.71 | 0.72 |
| TiC | 0.87 | 0.88 | 0.93 | 0.8 | 0.9 | 0.9 | 0.62 | 0.91 | 0.85 |
| Ge | 0.96 | 0.98 | 0.87 | 0.1 | 0.98 | 0.96 | 0.22 | 0.98 | 0.76 |
| Si | 0.99 | 1 | 0.94 | 1 | 1 | 1 | 1 | 0.99 | 0.99 |
| Fe | 0.91 | 0.9 | 0.92 | 0.33 | 0.86 | 0.83 | 0.16 | 0.92 | 0.73 |
| Ta | 0.99 | 1 | 0.96 | 0 | 0.99 | 0.99 | 0.86 | 0.99 | 0.85 |
| W | 0.99 | 1 | 0.78 | 0.46 | 0.99 | 0.98 | 0.73 | 1 | 0.86 |
| Al4CoNi2 | 0.86 | 0.99 | 0.99 | 0.95 | 0.99 | 0.99 | 0.38 | 0.98 | 0.89 |

**Figure 10:** Performance of MCD on predicting space group types of the chemically disordered equivalent structure when fluorite-type structures are removed from simulated training dataset. Scores from 8 different runs are recorded. a) With all phases included in experimental data. b) With patterns from NiAl and Al removed.

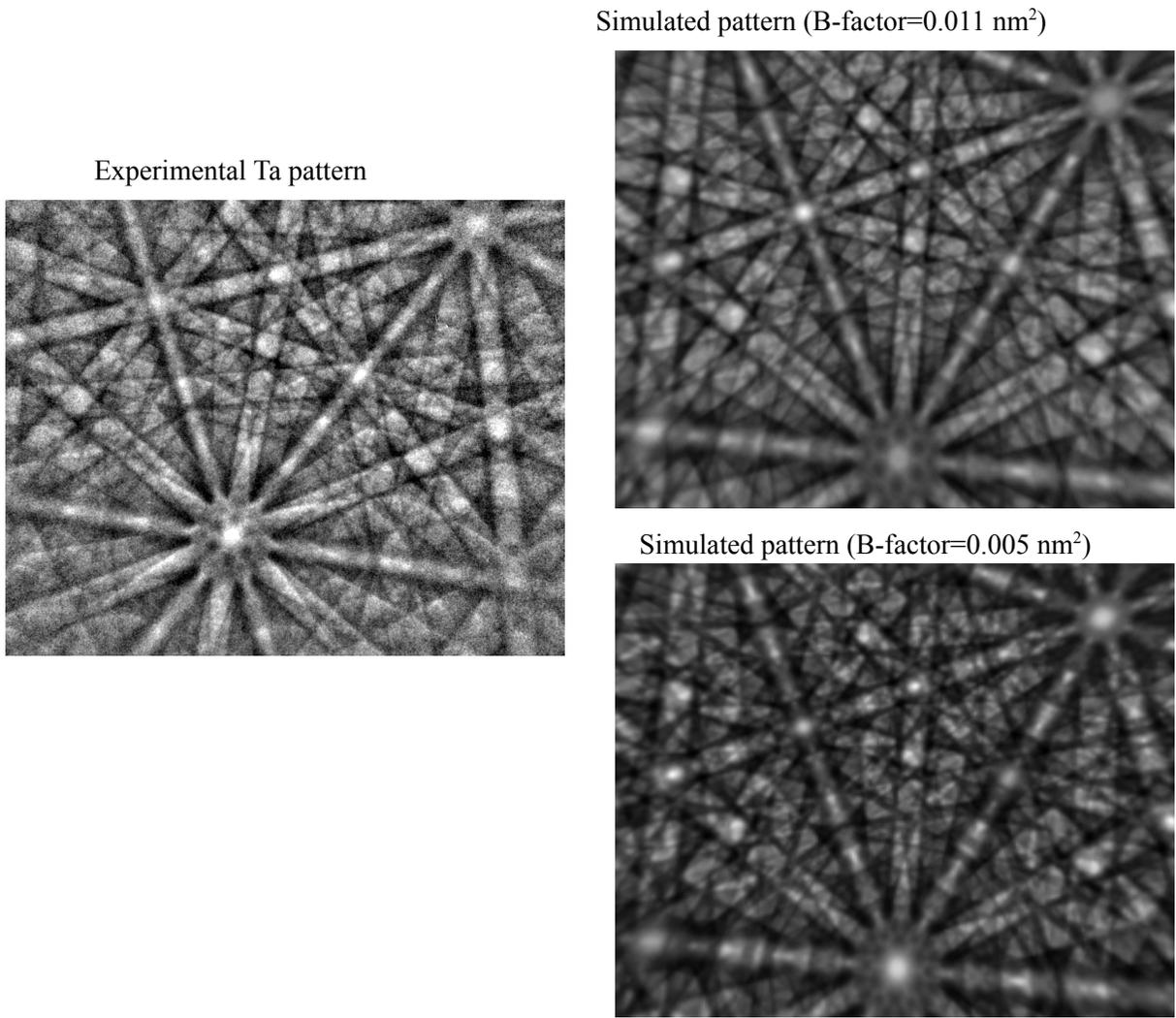

**Figure 11:** An experimental pattern with corresponding simulated patterns of Ta at similar orientations with different B-factors (also referred to as "Debye-Waller factors").